\documentclass[preprint,review,12pt]{elsarticle}

\usepackage{amsmath}

\usepackage{amssymb}
\usepackage{amsmath}
\usepackage{hyperref}

\usepackage{acro}

\DeclareAcronym{EDMD}{
  short=EDMD,
  long=Event-Driven Molecular Dynamics
}
\DeclareAcronym{CFD}{
  short=CFD,
  long=Computational Fluid Dynamics
}
\DeclareAcronym{DFT}{
  short=DFT,
  long=Density Functional Theory
}
\DeclareAcronym{PTV}{
  short=PTV,
  long=Particle Tracking Velocimetry
}
\DeclareAcronym{RH}{
  short=RH,
  long=Relative Humidity
}
\DeclareAcronym{DAQ}{
  short=DAQ,
  long=Data AcQuisition
}

\DeclareAcronym{FIBCs}{
  short=FIBCs,
  long=Flexible Intermediate Bulk Containers
}

\DeclareAcronym{SEM}{
  short=SEM,
  long=Scanning Electron Microscopy
}

\journal{Journal of Electrostatics}

\begin{document}

\begin{frontmatter}



\title{The Scaling of Triboelectric Charging Powder Drops for Industrial Applications} 

\author[Bristol]{Tom F. O'Hara\corref{cor1}}
\cortext[cor1]{Corresponding author: Tom F. O'Hara, Email: tom.ohara@bristol.ac.uk}
\author[Syngenta]{Ellen Player}
\author[Syngenta]{Graham Ackroyd}
\author[Syngenta]{Peter J. Caine}
\author[Bristol]{Karen L. Aplin}

\affiliation[Bristol]{organization={University of Bristol, School of Civil, Aerospace and Design Engineering},
            addressline={University Walk}, 
            city={Bristol},
            postcode={BS8 1TR}, 
            country={United Kingdom}}

\affiliation[Syngenta]{organization={Syngenta, Process Hazards},
            addressline={509 Leeds Road}, 
            city={Huddersfield},
            postcode={HD2 1XX}, 
            country={United Kingdom}}

\begin{abstract}
Triboelectrification of granular materials is a poorly understood phenomenon that alters particle behaviour, impacting industrial processes such as bulk powder handling and conveying. At small scales ($<$ 1 g) net charging of powders has been shown to vary linearly with the total particle surface area and hence mass for a given size distribution. This work investigates the scaling relation of granular triboelectric charging, with small, medium ($<$ 200~g), and large-scale ($\sim$ 400~kg) laboratory testing of industrially relevant materials using a custom powder dropping apparatus and Faraday cup measurements. Our results demonstrate that this scaling is broken before industrially relevant scales are reached. Charge (Q) scaling with mass (m) was fitted with a function of the form $Q \propto m^b$ and $b$ exponents ranging from $0.68\ \pm\ 0.01$ to $0.86\ \pm\ 0.02$ were determined. These exponents lie between those that would be expected from the surface area of the bulk powder ($b = 2 / 3$) and the total particle surface area ($b = 1$). This scaling relation is found to hold across the powders tested and at varying humidities.
\end{abstract}

\begin{graphicalabstract}
\includegraphics[width=1.0\textwidth]{Graphical_Abstract.pdf}
\end{graphicalabstract}

\begin{highlights}
\item Linear scaling of charging in bulk powder handling breaks down at larger powder masses.
\item Bulk powder surface area predicts charging better than total particle surface area.
\item Materials with humidity-dependent resistivity show variable charging but follow common scaling.
\end{highlights}


\end{frontmatter}



\section{Introduction}

A wide range of industrial processes are highly dependent on granular materials, many of which can become highly charged through poorly understood frictional (triboelectric) charging mechanisms \cite{matsusaka_triboelectric_2010, lacks_long-standing_2019}. Charge accumulation can disrupt normal material flow patterns, cause unwanted adhesion, reduce yields, and even cause dust cloud explosions through electrostatic discharges \cite{pingali_use_2009, glor_electrostatic_2005}. Electrostatic particle charges are known to accumulate through industrial processes such as pneumatic conveying, sieving, and fluidised bed operations \cite{wilms_ml_2024, deng_electrostatic_2023, murtomaa_electrostatic_2003}.

The extent of particle charging and the ability to retain this charge are highly dependent on both material properties and environmental factors. Material properties known to affect charging include conductivity, surface roughness, wettability, and particle size, among others \cite{heinert_decay_2022, jantac_triboelectric_2025, grunebeck_size_2024}. Environmental conditions such as temperature and relative humidity can also have a significant influence on tribocharging, with complex and interplaying effects\cite{xu_experimental_2023, cruise_triboelectric_2023}.


The inherent unpredictability of triboelectric charging in industrial processes complicates efforts to prevent electrostatic discharges, which may serve as ignition sources for dust cloud explosions \cite{puttick_avoidance_2008}. Consequently, oxygen removal through nitrogen inertisation remains the primary safety measure in many industrial processes, despite the significant environmental and economic costs associated with energy-intensive nitrogen separation \cite{choi_experimental_2015, aneke_potential_2015}. One area of particular concern is the transfer of powder into or from reactor vessels \cite{pey_charging_2022}. This transfer often involves \ac{FIBCs}, which vary in material composition and are used with diverse transfer systems \cite{glor_electrostatic_2005, ebadat_testing_1996}. While gravity-fed operations are preferred for their low environmental and economic cost, the extent of charging in these procedures is poorly understood. Currently, gravity-fed powder drops are typically limited to 3 m in height to minimise the perceived risk from charging, despite limited experimental evidence supporting this guideline \cite{puttick_avoidance_2008}. Some operations divide larger drops into multiple smaller drops to comply with this rule, though the effect on overall charging remains unclear. This work aims to address the need for a better understanding of tribocharging in powder drops and similar processes.

Understanding the relationship between charging and total powder mass is crucial for industrial scale-up. Small-scale studies (using masses of a few grams or less) demonstrated a linear relationship between net triboelectric charging and total particle surface area \cite{zarrebini_tribo-electrification_2013, ohara_faraday_2025}. This linearity is supported by \ac{CFD} and hard particle dynamics simulations, which predict that this relationship holds regardless of particle concentration and packing density \cite{alfano_computational_2021, lacks_effect_2007}. For a constant size-distribution, since total surface area varies linearly with mass, net triboelectric charging would also theoretically scale linearly with mass. However, validating these assumptions up to industrial scales remains a critical challenge that requires experimental investigation.

This study investigates charge scaling across multiple mass ranges. First, we examine small-scale drops (mass $<$ 2~g) in Section \ref{Small-Scale Measurement} to verify the linear charge-mass relationships reported in previous works \cite{zarrebini_tribo-electrification_2013, alfano_computational_2021}. Additional SEM analysis, particle-size characterisation, and powder resistivity measurements provide material insights. We extend the investigation to medium-scale tests (1-200 g) in Section \ref{Medium and large-Scale Measurement} using polypropylene, flour, and wettable sulfur, and scale up to industrial-level masses (400 kg) with polypropylene, typical of an industrial powder handling process. Finally, the influence of \ac{RH} on resistivity is measured to examine the effect on the charge scaling relation in Section \ref{Relative Humidity Resistivity Influence}.

\section{Methodology}

\begin{figure}[htbp]
    \centering
    \includegraphics[width=1.0\textwidth]{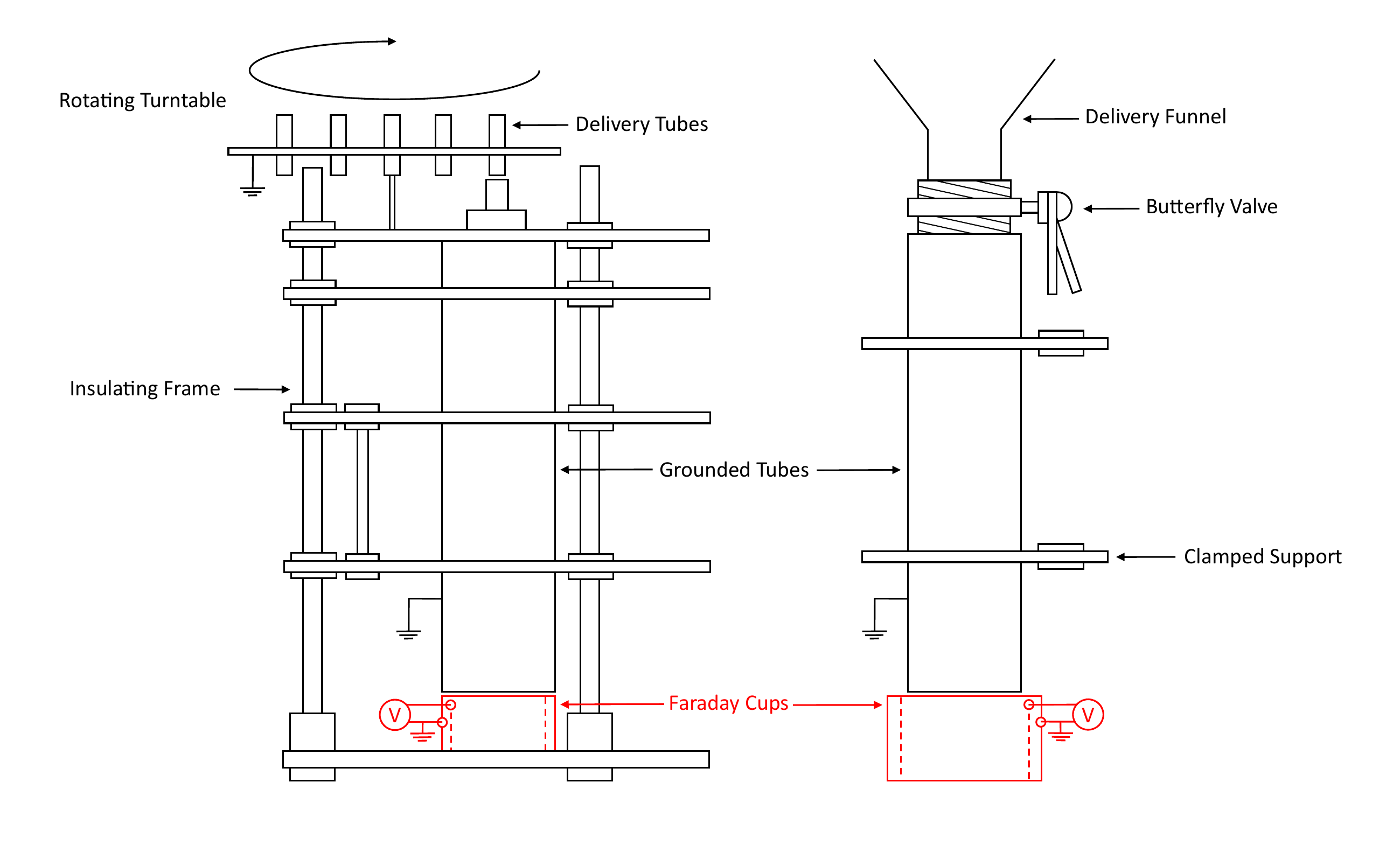}
    \caption{The experimental setup schematic for the charge drop apparatus used for small-scale 0.37~m drops (left) and medium-scale 0.50~m (right) powder drops. The small-scale apparatus is adapted from previous work \cite{ohara_faraday_2025, houghton_triboelectric_2013}}
    \label{Medium_Scale_Rig_Setup}
\end{figure}

To investigate the charge scaling with mass, three experimental setups of different scales were employed. Schematics of the small-scale ($<$ 2~g) and medium-scale ($<$ 200~g) setups can be seen in Figure \ref{Medium_Scale_Rig_Setup}. The small and medium-scale setups are similar, except for their size and a minor change in the release mechanism, from a stainless steel sliding trap door to a stainless steel butterfly valve. The large-scale measurements ($\sim$ 400~kg) were conducted by dropping powder from an FIBC, into a stainless steel hopper using a forklift truck to raise the FIBC. There are some physical limitations of handling this scale of material, such as having to use an FIBC with a conductive lining, rather than an entirely grounded stainless steel container as used for the other scale experiments.

The primary measurement technique utilised Faraday cups for measuring changes in voltage ($\Delta V$), which were converted to changes in charge ($\Delta Q$) using the known capacitance ($C$) of the system following $\Delta V = \Delta Q / C$. The capacitance values were determined to be 130~pF, 154~pF, and 19.7~nF for the small, medium, and large-scale setups, respectively. These values were obtained by applying a constantly changing voltage $dV / dt$ and determining the capacitance from the measured current ($I$) using the relation $I = C \cdot dV / dt$ \cite{aplin_self-calibrating_2001, houghton_triboelectric_2013, ohara_faraday_2025}.

 For measurement of voltage, the Faraday cups were connected by a triaxial BNC to a \href{https://ilg.physics.ucsb.edu/Courses/RemoteLabs/docs/Keithley6514manual.pdf}{Keithley 6514 system electrometer}. To minimise noise in the small-scale measurements, short and rigid connections were utilised; however, for the larger measurements, flexible cables could be used as the signal-to-noise ratio increased \cite{keithley_low_1998}. The electrometer was connected to a \href{https://www.ni.com/docs/en-US/bundle/usb-6211-specs/page/specs.html}{USB-6210} Data Acquisition (DAQ) device from National Instruments operated using  \href{https://www.ni.com/en/support/downloads/software-products/download.labview.html}{LabVIEW$^{\circledR}$ 2024-Q1} to record the data. The DAQ was also plugged into a \href{https://datasheet.octopart.com/386-Adafruit-Industries-datasheet-81453130.pdf}{DHT11 humidity and temperature sensor} using an \href{https://docs.arduino.cc/resources/datasheets/A000067-datasheet.pdf}{Arduino Mega 2560} to record the relative humidity and temperature. The temperature and humidity were not controlled in most experiments, but were recorded. Some of the small-scale experiments were conducted in a humidity-controlled room that operated using a Ebac DD400 desiccant dehumidifier and a Mitsubishi MSZ-AP25VGK air conditioning unit. The variations in environmental conditions are outlined in Section \ref{Results and Discussion}. The post-processing fittings were carried out using a Levenberg-Marquardt algorithm implemented by \href{https://docs.scipy.org/doc/scipy/reference/generated/scipy.optimize.minimize.html}{scipy.optimize.minimize} package for \href{https://docs.python.org/3/}{Python 3.12.2}.

The \ac{SEM} images in this work were produced using a Hitachi TM3030Plus Tabletop Microscope. The particle sizing was carried out using the Malvern Mastersizer 3000 with aero-dispersion. Powder resistivity measurements were conducted using an Agilent 4339B High Resistance Meter connected to a powder resistivity cell. The cell had dimensions of 10~cm $\times$ 1~cm (area, $A$) with a sample length ($L$) of 1~cm. Using an applied voltage ($V$) of up to 1000~V and measuring the resulting current ($I$), the resistivity ($\rho$) was calculated according to $\rho = (V/I) \cdot (A/L)$.

\section{Results and Discussion}
\label{Results and Discussion}
\subsection{Small-Scale Measurement}
\label{Small-Scale Measurement}

\begin{figure*}[htbp]
    \centering
    \includegraphics[width=1.0\textwidth]{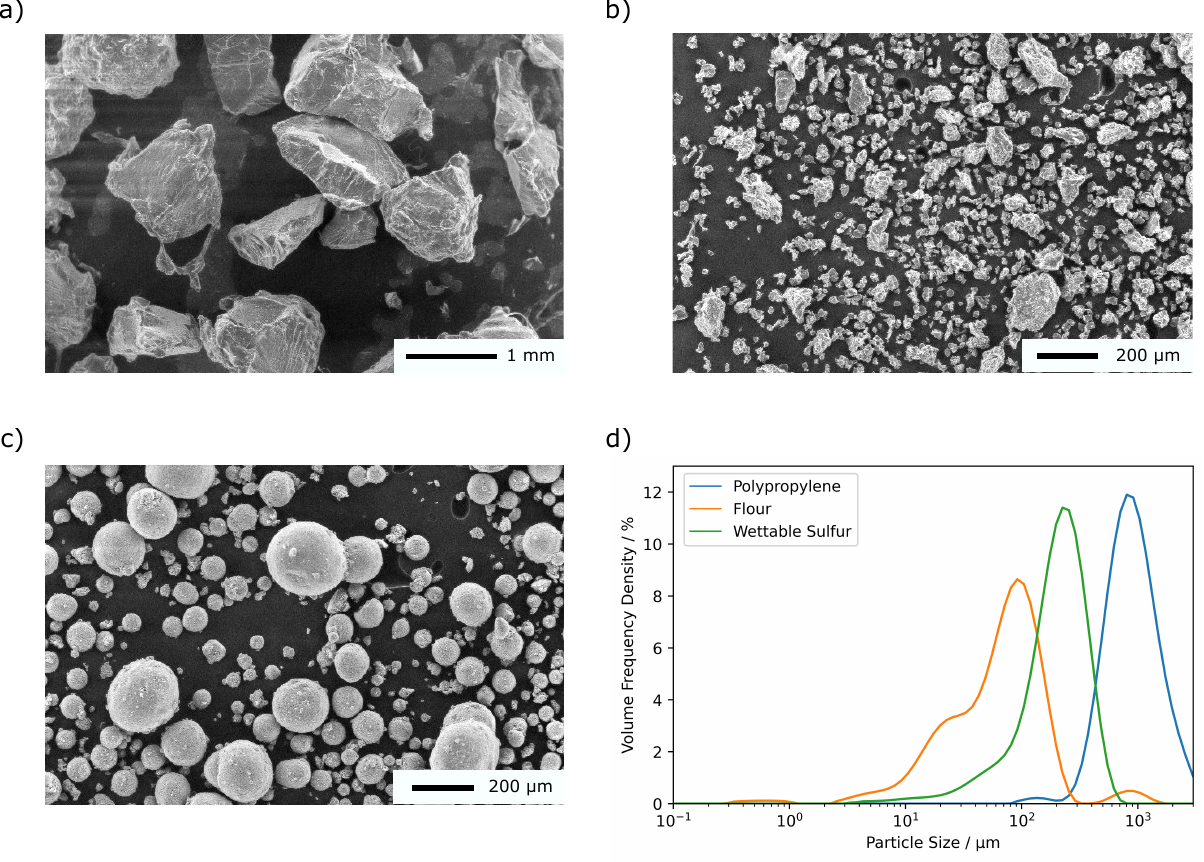}
    \caption{\ac{SEM} images of a) polypropylene, b) flour, and c) wettable sulfur, with their respective scale-bars shown. d) The Mastersizer 3000 sizing data of the air-dispersed particle sizes by volume frequency density.}
    \label{SEM_and_sizing}
\end{figure*}

To investigate mass-scaling for industrially relevant powder charging, three representative materials with distinct properties were selected: polypropylene pellets, wheat flour, and wettable sulfur. Polypropylene, being highly insulating, features large rough pellets with a modal particle size of 800~$\mu$m (Figure \ref{SEM_and_sizing}). Wheat flour, composed primarily of complex carbohydrates, consists of irregularly shaped particles with abundant surface hydroxyl groups capable of electron transfer during contact-separation events. Wettable sulfur, a common fungicide, exhibits a well-defined colloidal structure (Figure \ref{SEM_and_sizing}c) resulting from its spray drying manufacturing process, and demonstrates distinct hydrophilic behaviour. However, the diversity in particle size, morphology, surface properties, and flowability strengthens the robustness and generalizability of this study.

\begin{figure}[htbp]
    \centering
    \includegraphics[width=0.6\textwidth]{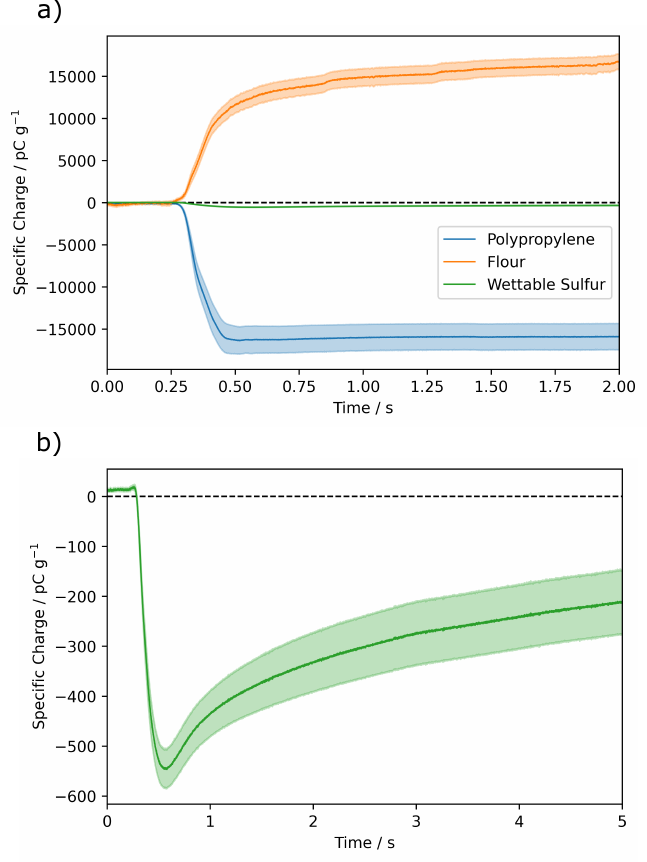}
    \caption{The experimental Faraday cup traces for small-scale drops of powder into the cup for a) the three powders investigated and b) wettable sulfur only. Each trace is an average of around 50 powder drop traces, with the error regions representing one standard error on the mean at each time step.}
    \label{small_scale_traces}
\end{figure}

Powder drops were carried out in a fume hood, with at least 50 samples of each powder dropped and measured at a range of masses so that an average can be taken of their specific charge (charge divided by mass). The diverse charging behaviours of the three investigated materials can be seen in Figure \ref{small_scale_traces}a, whereby the polypropylene and flour both charge, oppositely, to a similar large magnitude. The wettable sulfur charges much less and also has a differently shaped response, Figure \ref{small_scale_traces}b. The shape of the wettable sulfur trace shows a negative dip before rising again as positive particles land in the Faraday cup. The shape of the trace produced suggests that a large proportion of its charge has arisen from the same material particle-particle contacts between wettable sulfur particles\cite{ohara_faraday_2025}. Whereas for the polypropylene and flour, the majority of the charging is expected to arise from different-material interactions, such as those between particles and the stainless-steel walls of the release chamber. It is possible to estimate the extent of charging arising from particle-particle or particle-wall interactions in more detail with the use of a model\cite{ohara_faraday_2025}. However, as the aim of this work is to look at the total charging, the key information that is extracted from each trace is either the difference between the start and end of the trace (referred to as total charging) and the difference between the maximum and minimum within each trace (referred to as charge magnitude).

\begin{figure}[htbp]
    \centering
    \includegraphics[width=0.6\textwidth]{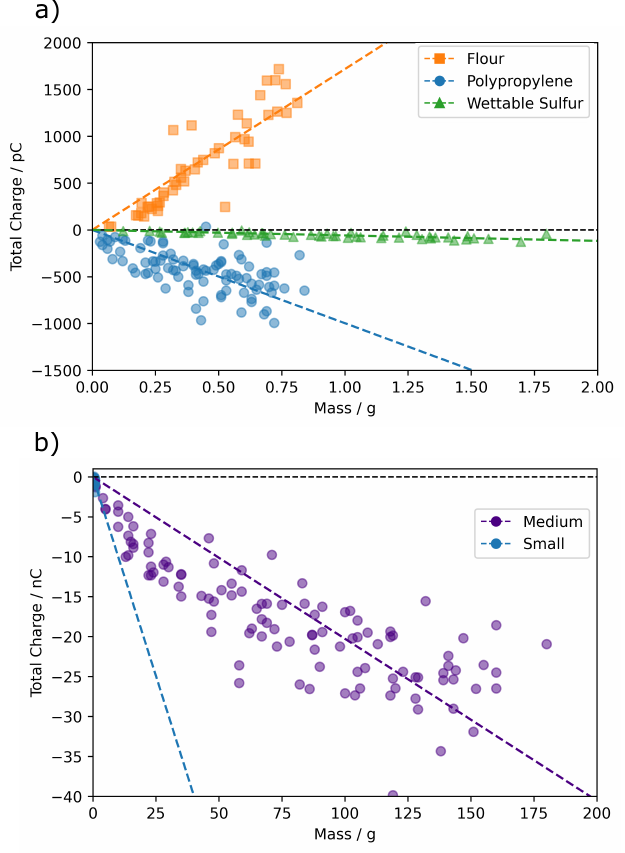}
    \caption{a) The total charging from small-scale Faraday cup experiments of powders (coloured points) is plotted against mass, with linear fits shown as coloured lines. b) Comparing the linear fits of medium and small-scale experiments with polypropylene demonstrates that the same linear relationship does not hold across orders of magnitude in mass.}
    \label{small_scale_linearity}
\end{figure}

The small-scale powder drops partially demonstrate the expected approximately linear charging relation with mass, as seen in Figure \ref{small_scale_linearity}. There is also a reasonably large amount of scatter, which will partially be due to the stochastic nature of triboelectric charging generally and also fluctuations in environmental conditions and small changes in handling. In the lab, the \ac{RH} is not controlled, but fluctuates around $37 \pm 10$\% and the temperature is 21 $\pm$ 4 $^\circ\text{C}$, where the error shown is two standard deviations. Overall, the $R^2$ values in Figure \ref{small_scale_linearity}a for the flour, polypropylene, and wettable sulfur are: 0.78, 0.37, and 0.59, demonstrating that whilst the relationship between total charging and mass is somewhat linear at low masses, the correlation is still weak. The same experiment was also carried out in a controlled humidity room for a subset of the polypropylene, where \ac{RH} is $24 \pm 2$\% and the temperature is 21.7 $\pm$ 0.8 $^\circ\text{C}$. The low-humidity room experiments for small-scale polypropylene have a KS statistic of 0.19 and a p-value of 0.90, when compared with those not in the low-humidity room. This indicates that there is no statistical evidence for differences between distributions. This suggests that humidity variations did not significantly affect the results for polypropylene.

Extrapolating the linearity from small to medium-scale experiments overestimates the charging, as shown for polypropylene in Figure \ref{small_scale_linearity}b. This overestimate also holds for extrapolating linear fits for flour and wettable sulfur. Furthermore, as well as being inconsistent across the mass scales, linear fits of the medium-scale drops have even less favourable $R^2$ values of: 0.40, 0.40, and 0.19. Therefore, a more accurate scaling relation is required.

\subsection{Medium and large-Scale Measurement}
\label{Medium and large-Scale Measurement}

\begin{figure*}[htbp]
    \centering
    \includegraphics[width=1.0\textwidth]{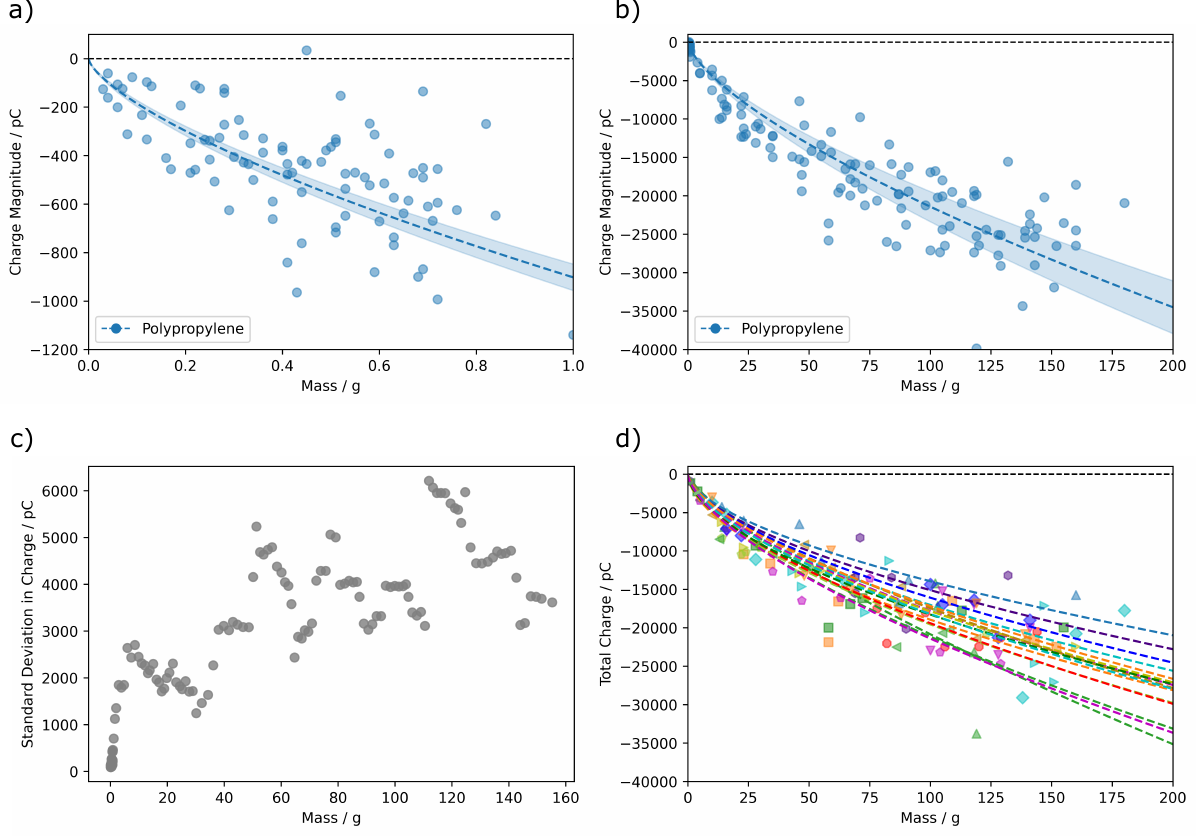}
    \caption{The total particle charging extracted from Faraday cup traces of polypropylene at a) small and b) medium-scale. The power law fit is: $y = ax ^b$ is shown where $a = -900 \pm 30$, $b= 0.69 \pm 0.01$, and the $R^2$ value is 0.91. The error region is shaded for 2 standard errors in the fitting parameters. c) Standard deviation in charge ($\sigma_Q$) versus mass, calculated using a sliding window of 10 data points, showing increasing charge variability with mass. d) Power-law fits for medium-scale polypropylene drop subsamples, with each subset's data and corresponding fit line shown in matching colours.}
    \label{medium_scale_fit}
\end{figure*}

As stated in Section \ref{small_scale_linearity}, a linear relationship is expected between the total charging and mass if the charging is proportional to the total particle surface area and the particle size distribution is maintained. However, in bulk powder handling in which particle-wall interactions dominate the charging, the area available for triboelectric charging may be more closely correlated with the outer area of the bulk powder, as this is where the particle-wall interactions can take place. This effective area scaling is geometry dependent, but for a simplified cube or sphere, the surface area ($A$) varies with the length or diameter ($L$) as $A \propto L^2$ and as the mass ($m$) is related by $m \propto L^3$, then $A \propto m^{2/3}$. This would predict that the charging may scale more with a power law of $2/3$, rather than the effective power of $1$ in a linear fit. To investigate this, the total charging ($Q$) can be fitted to a power-law function of the form: $Q = am^b$, where $a$ and $b$ are fitting parameters. It can be assumed that $Q = 0$ at $m = 0$, hence, no intercept term is included. For polypropylene, this fit finds $b= 0.69 \pm 0.01$ ($R^2 = 0.91$), which is in close agreement with the $2/3$ value predicted by a bulk powder surface relation and is consistent across the mass scales, as seen in Figure \ref{medium_scale_fit}a and b.

As multiple orders of magnitude are compared and the spread of total charge increases with mass, a weighted fit is required to prevent extreme values from dominating the result, as illustrated in Appendix Figure \ref{bad_fittings}a. This work employs a sliding window variance approach \cite{zhang_variance_2007}. By sorting the data by mass and scanning over windows of fixed width ($w = 10$ data points in this work), the mean mass ($\mu_m$) and mean total charge ($\mu_Q$) for each window $k$ are calculated as:
\begin{equation*}
\begin{array}{c@{\hskip 1cm}c@{\hskip 1cm}c}
\displaystyle \mu_m(k) = \frac{1}{w} \sum_{i=k}^{k+w-1} m_i &
\text{and} &
\displaystyle \mu_Q(k) = \frac{1}{w} \sum_{i=k}^{k+w-1} Q_i\text{,}
\end{array}
\end{equation*}
from which the local standard deviation of charge, $\sigma_Q$, is determined as:
\begin{equation*}
\sigma_Q(k) = \sqrt{\frac{1}{w} \sum_{i=k}^{k+w-1}(Q_i - \mu_Q(k))^2}\text{.}
\label{charge_spread}
\end{equation*}
The sliding window standard deviation associated with each mass value is then estimated by linear interpolation, and the corresponding weight of $1/\sigma_Q$ is applied in the least-squares fit.

To mitigate the effects of fluctuations in environmental conditions, the experiments are carried out in a random order concerning mass to avoid any systematic bias in environmental variables. Subsets of the medium-scale data that were carried out within a short space of time are also fitted with the small-scale data, as seen in Figure \ref{medium_scale_fit}d, giving a range of power-law fits averaging $b = 0.62 \pm 0.04$, which also agrees with the overall exponent of $b= 0.69 \pm 0.01$. This analysis not only provides additional statistical evidence for the power-law fitting but also yields a larger uncertainty that may be more physically realistic than the error from the overall fitting parameters, as it accounts for variations across subsets.

\begin{figure}[htbp]
    \centering
    \includegraphics[width=0.6\textwidth]{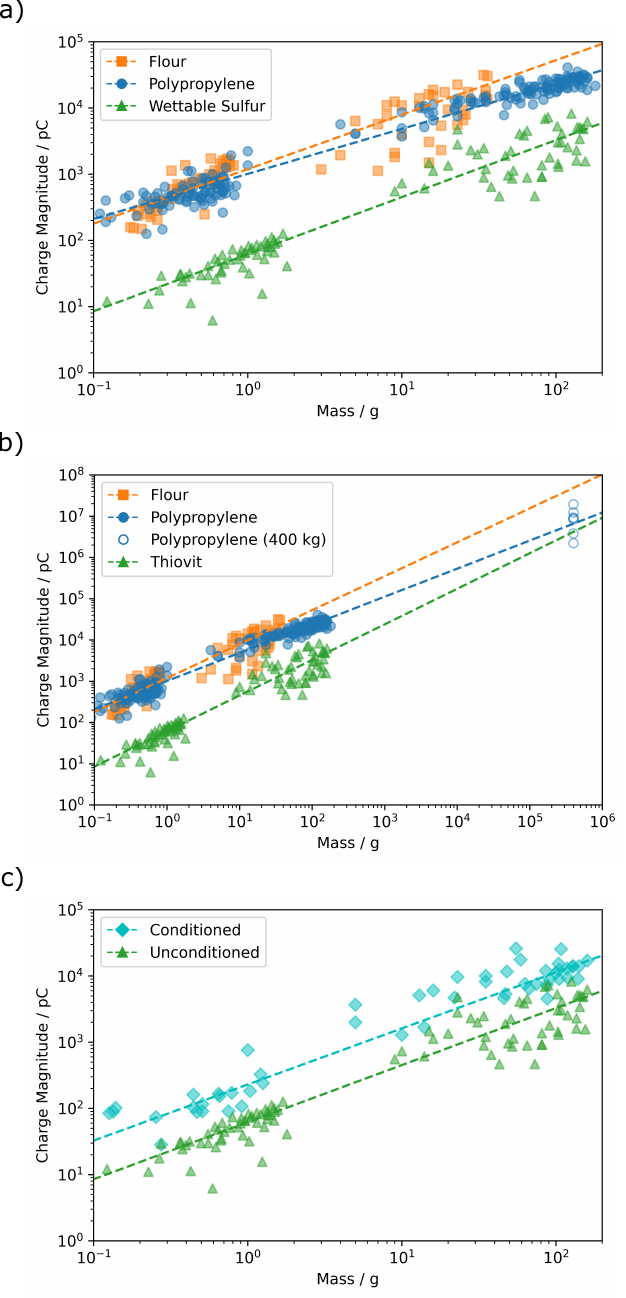}
    \caption{a) Log-log plots of the total particle charging extracted from Faraday cup traces of polypropylene flour and wettable sulfur, expressed as charge magnitude to enable logarithmic analysis. b) The same fittings as in a) but with large-scale polypropylene data also displayed. c) Comparison of the wettable sulfur without conditioning and at 20\% RH and 20~$^\circ$C and without conditioning.}
    \label{log_log_plots}
\end{figure}

The fits for flour and wettable sulfur are included in the log-log plot shown in Figure \ref{log_log_plots}a with power law exponents of $b= 0.82 \pm 0.02$ and $b= 0.86 \pm 0.02$, respectively. These powders deviated from the $2/3$ predicted by the bulk surface area but are still less than the value of $1$ predicted by the total particle surface area. This suggests that a generalised scaling of $Q = am^b$, where $2/3 \leq b \leq 1$ may be broadly appropriate for the triboelectric charging of powders. The large-scale drops at 400~kg are also included in \ref{log_log_plots}a, and the large-scale experiments are consistent with the fits determined independently from the small and medium-scale experiments. These results demonstrate that the power law relationships obtained are applicable across many orders of magnitude.

\subsection{Relative Humidity Resistivity Influence}
\label{Relative Humidity Resistivity Influence}

In Figure \ref{log_log_plots}a it is also notable that variability in the flour and wettable sulfur data is greater than that of polypropylene. This may be partly explained by variation in the powders' resistivities, shown in Table \ref{tab:resistivity}. The measured resistivities are lower than may be expected due to contamination of the samples, as these measurements were taken after conducting the charge tests on the powders. However, this makes the powder resistivities more representative of those of the powders used in the experiments. The low unconditioned resistivity of wettable sulfur (at $37 \pm 5$\% RH and 21~$\pm$~2~$^\circ\text{C}$) partially explains the low charging observed relative to the other powders investigated. The key difference between the powder resistivities is that polypropylene changes much less after conditioning, particularly in comparison to the wettable sulfur. This indicates that its resistivity is less affected by fluctuations in \ac{RH}, and is perhaps less susceptible to variability from fluctuations in environmental conditions.

\begin{table}[h]
\centering
\caption{The resistivity ($\rho$) values of materials conditioned at 20\% RH and 20~$^\circ$C for 24 hours, compared to unconditioned samples at $37 \pm 10$\% RH and 21 $\pm$ 4 $^\circ\text{C}$.}
\begin{tabular}{|l|c|c|}
\hline
Material & $\rho$ / $\Omega\cdot$m & Unconditioned $\rho$ / $\Omega\cdot$m \\
\hline
Polypropylene      & $8.2 \times 10^{12}$ & $1.0 \times 10^{12}$ \\
Flour              & $3.8 \times 10^{10}$ & $2.3 \times 10^{9}$ \\
Wettable Sulfur    & $1.1 \times 10^{11}$ & $1.2 \times 10^{7}$ \\
\hline
\end{tabular}
\label{tab:resistivity}
\end{table}

The reduced resistivity of wettable sulfur leads to reduced charging, as demonstrated in Figure \ref{log_log_plots}c. However, the power law fit is preserved ($b = 0.86 \pm 0.02$ to $b = 0.85 \pm 0.03$ with conditioning) despite a large increase in charge magnitude ($a = 62 \pm 3$ pC g$^{-1}$ to $a = 230 \pm 30$ pC g$^{-1}$). This gives evidence that the power-law scaling with $2/3 \leq b \leq 1$ is conserved despite changes in environmental conditions.

\section{Conclusions and Outlook}
Understanding how triboelectric charging scales with mass is crucial for translating laboratory findings to industrial processes, yet studies have been largely limited to small-scale ($<$ 1 g) measurements. This work systematically investigated charge scaling across multiple mass ranges (from grams to hundreds of kilograms) using three representative materials with distinct properties, combining Faraday cup measurements with careful environmental monitoring and statistical analysis to bridge this knowledge gap.

The linear scaling of charging measured with small (g) samples of powder is shown not to apply at larger masses, above gram or kilograms. A new scaling closer to the surface area of the bulk powder, rather than the total particle surface area, is found in three materials with differing properties and charging characteristics. It is proposed that a power-law scaling for total charge $Q$) against mass ($m$) of $Q = am^b$, where $2/3 \leq b \leq 1$ may be broadly generalisable for the triboelectric charging of powders. Charge predictions are found to agree across multiple orders of magnitude in powder mass, using a sliding window variance approach, despite some variation in the environment of the measurements.

This research has the capability to inform policy decisions around the safe processing of bulk powders and increase the opportunity for process design to better adapt to powder charging risks and losses. Further work is required to evaluate whether the $Q = am^b$ scaling relation is universal across a broader range of powders. Systematic data collection, standardised powder handling and increased control of environmental conditions could also provide areas for further insights, with reduced statistical uncertainty in the scaling of triboelectric charging.

\section*{CRediT authorship contribution statement}
\textbf{Tom F. O'Hara:} conceptualisation, methodology, formal analysis, validation, data curation, writing – original draft, writing – review and editing, project administration. \textbf{Ellen Player} supervision, writing – review and editing. \textbf{Graham Ackroyd:} supervision, resources. \textbf{P. J. Caine:} investigation. \textbf{Karen L. Aplin:} supervision, writing – review and editing, funding acquisition.

\section*{Data availability}
All the data used in this work are freely available through the Materials Data Facility \cite{ohara_scaling_2025}. The code used for post-processing analysis and plotting can be accessed under a GPL-3.0 public use license \cite{tom_ohara_vb22224industrial_powder_scaling_plotter_2025}.

\section*{Declaration of competing interest}
There are no conflicts to declare.

\section*{Acknowledgements}
The authors would like to thank Stephen Pearson at Syngenta for the constructive discussions around the data science of this work and the larger hazards team at Syngenta for their support and guidance regarding the use of equipment. Funding was provided by the Engineering and
Physical Sciences Research Council through the Centre for Doctoral Training in Aerosol Science (no. EP/S023593/1) and materials provided by the Process Hazards Department at Syngenta Huddersfield.


\appendix
\section{Fitting Validation}
\label{appendix}
\begin{figure*}[htbp]
    \centering
    \includegraphics[width=0.6\textwidth]{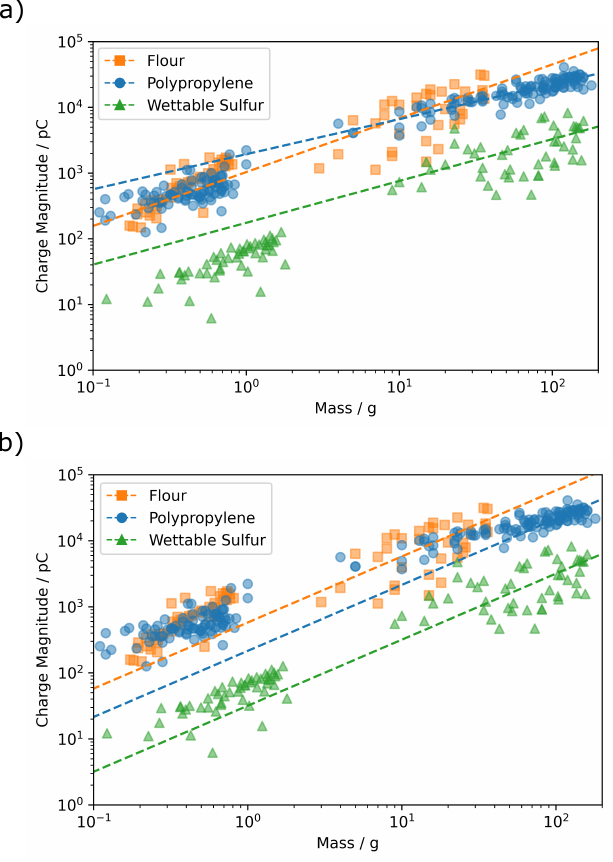}
    \caption{The log-log plots of the total particle charging extracted from Faraday cup traces of polypropylene flour and wettable sulfur, expressed as charge magnitude. The fittings are carried out a) with the form $y = a x^b$  without a weighted fitting for variance and b) with the form $y = mx$.}
    \label{bad_fittings}
\end{figure*}

The mass charge fittings in Figure \ref{bad_fittings} demonstrate that unweighted fits or those without inverse variance weighting fail to fit appropriately across multiple orders of magnitude of mass, unlike the more successful fits shown in Figure \ref{log_log_plots}.

\bibliographystyle{elsarticle-num-names} 
\bibliography{references}

\end{document}